\documentstyle[psfig,natbib]{elsart}

\begin{document}

\begin{frontmatter}
\title{Timing evolution of accreting strange stars}

\author[rostock,dubna]{D. Blaschke}
\author[pisa]{I. Bombaci}
\author[yerevan]{H. Grigorian\thanksref{DAAD}}
\author[yerevan]{G. Poghosyan\thanksref{DFG}}
\address[rostock]{Fachbereich Physik, Universit\"at Rostock, D-18051 Rostock,
Germany}
\address[dubna]{Bogoliubov  Laboratory of Theoretical Physics,\\
        Joint Institute for Nuclear Research, 141980, Dubna, Russia}
\address[pisa]{Dipartimento di Fisica ``E. Fermi'', Universit\'a di Pisa and
INFN Sezione di Pisa, 56127 Pisa, Italy}
\address[yerevan]{Department of Physics, Yerevan State University,
375025 Yerevan, Armenia}
\thanks[DFG]{Supported by DFG grant No 436 ARM 17/7/00}
\thanks[DAAD]{Supported by DAAD}

\begin{abstract}

It has been suggested that the QPO phenomenon in LMXB's could be
explained when the central compact object is a strange star. In
this work we investigate within a standard model for disk
accretion whether the observed clustering of spin frequencies in a
narrow band is in accordance with this hypothesis. We show that
frequency clustering occurs for accreting strange stars when
typical values of the parameters of magnetic field initial
strength and decay time, accretion rate are chosen. In contrast to
hybrid star accretion no mass clustering effect is found.

\noindent PACS number(s):  04.40.Dg, 12.38.Mh, 26.60.+c, 97.60.Gb
\begin{keyword}
accretion, accretion discs - stars: interiors - stars: magnetic fields -
pulsars: general - X-rays: binaries
\end{keyword}
\end{abstract}
\end{frontmatter}
\newpage

\section{Introduction}
The recent discovery of new phenomena in the physics of
X-ray binaries is giving a new and powerful tool to probe
dense hadronic matter 
\citep{MLP}
and possibly to infer the
existence of a deconfined phase of quark matter in the cores of the
central accretors in these systems.
Various signals of a phase transition to deconfined quark matter
have been suggested in the form of peculiar changes of observables,
such as the pulse timing 
\citep{gw00,accmag},
and the thermal evolution of isolated pulsars 
\citep{schaab,pagecool,cool}.
Even more intriguing than the existence of a quark core in a neutron star,
is the possible existence of a new family of compact stars
consisting completely of a deconfined mixture of {\it up} ({\it u}),
{\it down} ({\it d}), and {\it strange} ({\it s}) quarks, together with
an appropriate number of electrons to guarantee electrical neutrality.
In the literature, such compact stars have been referred to as
{\it strange quark stars} or shortly {\it strange stars} (SS),
and their constituent matter as {\it strange quark matter} (SQM).
In a series of recent papers 
\citep{dey+,strange}
it has been argued that the compact objects in some X-ray sources are likely
strange star candidates.
Particularly interesting candidates are the compact stars in the
newly discovered millisecond X-ray  pulsar SAX J1808.4--3658,
and in the atoll source 4U~1728--34.

In the present paper, we model the spin evolution of an accreting
strange star in a binary stellar system, from the so called
``death line'' up to the millisecond pulsar phase. We explore the
dependence of the SS spin evolution upon the mass accretion rate
and upon those physical quantities which regulate the temporal
evolution of the torque acting on the spinning star. We try to
constrain these quantities to have a population clustering of
strange stars in agreement with the spin frequency distribution
for observed Z sources in LMXBs with kHz QPOs. Possible signatures
to distinguish strange stars from ordinary neutron stars are also
briefly discussed.

\section{Accretion model and magnetic field evolution}

We consider the spin evolution of a strange star under mass accretion
from a low-mass companion star as a sequence of stationary states of
configurations (points) in the phase diagram spanned by the angular velocity 
$\Omega$ and the baryon number $N$.  
The process is governed by the change in angular momentum 
$J(N,\Omega)=I(N,\Omega)~\Omega$ 
\begin{equation}
\label{djdt}
\frac{d}{dt} (J(N,\Omega))= K_{\rm ext}~,
\end{equation}
where $I(N,\Omega)$ is the moment of inertia of the star and 
\begin{equation}
K_{\rm ext}= \sqrt{G M \dot M^2 r_0}- N_{\rm out}
\label{kex}
\end{equation}
is the external torque due to both the specific angular
momentum transferred by the accreting plasma and the magnetic plus
viscous stress given by $N_{\rm out}=\kappa \mu^2 r_c^{-3}$,
$\kappa=1/3$ 
\citep{lipunov}. 
 For a star with radius
$R$ and magnetic field strength $B$, the magnetic moment is given by
$\mu=R^3~B$. The co-rotating radius
$r_c=\left(GM/\Omega^2\right)^{1/3}$ is very large ($r_c\gg r_0$)
for slow rotators.
The inner radius of the accretion disc is
\[
r_0 \approx \left\{
\begin{array}{cc}
R~,&\mu < \mu_c \\
0.52~r_A~,&\mu \geq \mu_c
\end{array}
\right.
\]
where $\mu_c$ is that value of the magnetic moment of the star for
which the disc would touch the star surface.  The characteristic
Alfv\'en radius for spherical accretion with the rate $\dot M=m \dot N$
is $r_A=\left(2\mu^{-4} G M \dot M^2\right)^{-1/7}$ 
\citep{heuvel}.
Since we are
interested in the case of fast rotation for which the spin-up torque
due to the accreting plasma in Eq.  (\ref{kex}) is partly compensated
by $N_{\rm out}$, eventually leading to a saturation of the spin-up, we
neglect the spin-up torque in $N_{\rm out}$ which can be important only
for slow rotators 
\citep{gl}.

From Eqs.  (\ref{djdt}), (\ref{kex}) one can obtain a first order
differential equation for the evolution of angular velocity

\begin{equation}
\label{odoto}
\frac{d \Omega}{d t}=
\frac{K_{\rm ext}(N,\Omega)- K_{\rm int}(N,\Omega)}
{I(N,\Omega) + {\Omega}({\partial I(N,\Omega)}/{\partial \Omega})_{N}}~,
\end{equation}
where the internal torque term defined as 
\begin{equation}
\label{kint}
K_{\rm int}(N,\Omega)=\Omega~\dot N~
\left(\frac{\partial I(N,\Omega)}{\partial N}\right)_{\Omega}~.
\end{equation}

Solutions of (\ref{odoto}) are trajectories in the $\Omega - N$ plane
describing the spin evolution of accreting compact stars
. Since for the hybrid stars $I(N,\Omega)$ exhibits characteristic
functional dependences 
\citep{phdiag} 
at the deconfinement phase
transition line $N_{\rm crit}(\Omega)$ we expect observable
consequences in the $\dot P - P$ plane when this line is crossed. In the
case of SS we have no expectation of the mass clustering due to absence of
phase transition effects in strange matter.

In our model calculations we assume that both the mass accretion and
the angular momentum transfer processes are slow enough to justify the
assumption of quasistationary rigid rotation without convection.  
For a more detailed description of the method and analytic
results we refer to 
\citet{cgpb} 
and the works of 
\citet{thorne}, 
as well as 
\citet{chubarian}.

The time dependence of the baryon number for the constant accretion
rate $\dot N$ is given by
\begin{equation}
N(t)=N(t_0)+ (t-t_0)\dot N~.
\end{equation}
For the magnetic field of the accretors we consider the exponential
decay 
\citep{heuvel}
\begin{equation}
B(t)=[B(0) - B_{\infty}]\exp(-t/\tau_B)+ B_{\infty}~.
\end{equation}
We solve the equation for the spin-up evolution (\ref{odoto}) of the
accreting star for decay times $10^7\le \tau_B {\rm [yr]} \le 10^9$ and
initial magnetic fields in the range $0.2 \leq B(0){\rm [TG]}\leq 4.0
$.  The remnant magnetic field is chosen to be
$B_\infty=10^{-4}$TG\footnote[1]{1 TG= $10^{12}$ G} 
\citep{page}.

At high rotation frequency, both the angular momentum transfer from
accreting matter and the influence of magnetic fields can be small
enough to let the evolution of angular velocity be determined by the
dependence of the moment of inertia on the baryon number, i.e.  on the
total mass.  This case is similar to the one with negligible magnetic
field considered in 
\citep{shapiro,colpi,cgpb},
where $\mu \leq \mu_c$ in Eq.
(\ref{odoto}), so that only the so called internal torque term
(\ref{kint}) remains.

\section{Equation of state for strange quark matter}
To describe the properties of strange quark matter, we used
a recent model for the equation of state (EoS) derived by 
\citet{dey+}.
This model is based on a {\it dynamical} density-dependent approach
to confinement.
This EoS has asymptotic freedom built in, shows confinement
at zero baryon density, and deconfinement at high density.
In this model the quark interactions is described by a
color-Debye-screened interquark potential originating from gluon exchange,
and by a density-dependent scalar potential which restores chiral symmetry
at high density (in the limit of massless quarks).
This density-dependent scalar potential arises from the density dependence
of the in-medium effective quark masses $M_q$, which in the model of 
\citet{dey+} 
are taken to depend on the baryon number density $n_B$ according to
\begin{equation}
 M_q(n_B) = m_q  + 310~({\rm MeV})~\rm ~sech \Big(\nu
\frac{n_B}{n_0}\Big),
\end{equation}
where $q (= u,d,s)$ is the quark flavor index, $n_0 = 0.16$~fm$^{-3}$
is the normal nuclear matter density, and $\nu$ is a parameter.
The effective quark mass $M_q(n_B)$ goes from its constituent mass value
at zero density, to its current mass $m_q$ as $n_B$ goes to infinity.
Here we consider two different parameterizations of the 
EoS, which correspond to a different choice for the parameter $\nu$.
The  equation of state SS1 (SS2) corresponds to $\nu = 0.333$ ($\nu = 0.286$).
These two models for the EoS give absolutely stable SQM according to the
strange matter hypothesis 
\citep{bodmer,witten},
see Tab. 1.

Medium dependent mechanisms for confinement and their consequences
for the EoS of quark matter, have been explored by many authors
using different QCD motivated phenomenological models 
\citep{bkt90,b+99,bt,drago}.
In addition to the previous model for the EoS, we make use of the
MIT bag model EoS 
\citep{faja}
for strange quark matter
for non-interacting quarks, with strange quark mass $m_s = 150$~MeV,
massless {\it u}  and {\it d} quarks, and with bag constant
$B = 60$~MeV/fm$^3$ (hereafter the B60$_{150}$ EoS).

\begin{table}
\caption{Ground state properties of SQM for the equations of state used
in the present work.
$(E/A)_{gs}$ (in MeV) is the energy per baryon,
$\rho_{gs}$ ($\times 10^{14}$ g/cm$^3$) the mass density, and
$n_{gs}$ (fm$^{-3}$)  the baryon number density.
The two equations of state SS1 and SS2 differ for the choice
of the parameter $\nu$ entering in the expression of the in-medium quark
masses.}
\begin{center}
\renewcommand{\arraystretch}{1.4}
\setlength\tabcolsep{5pt}
\begin{tabular}{llll}
\hline\noalign{\smallskip}
         EoS      & $(E/A)_{gs}$ & $\rho_{gs}$ &$n_{gs}$ \\
\noalign{\smallskip}
\hline
\noalign{\smallskip}
       SS1          &    ~~~888    & 12.3    & 0.779  \\
       SS2          &    ~~~926    & 14.1    & 0.858  \\
       B60$_{150}$  &    ~~~836    & ~4.6    & 0.295  \\
\hline
\end{tabular}
\end{center}
\end{table}

\section{Results and discussion}

\begin{figure}[htb]
\psfig{figure=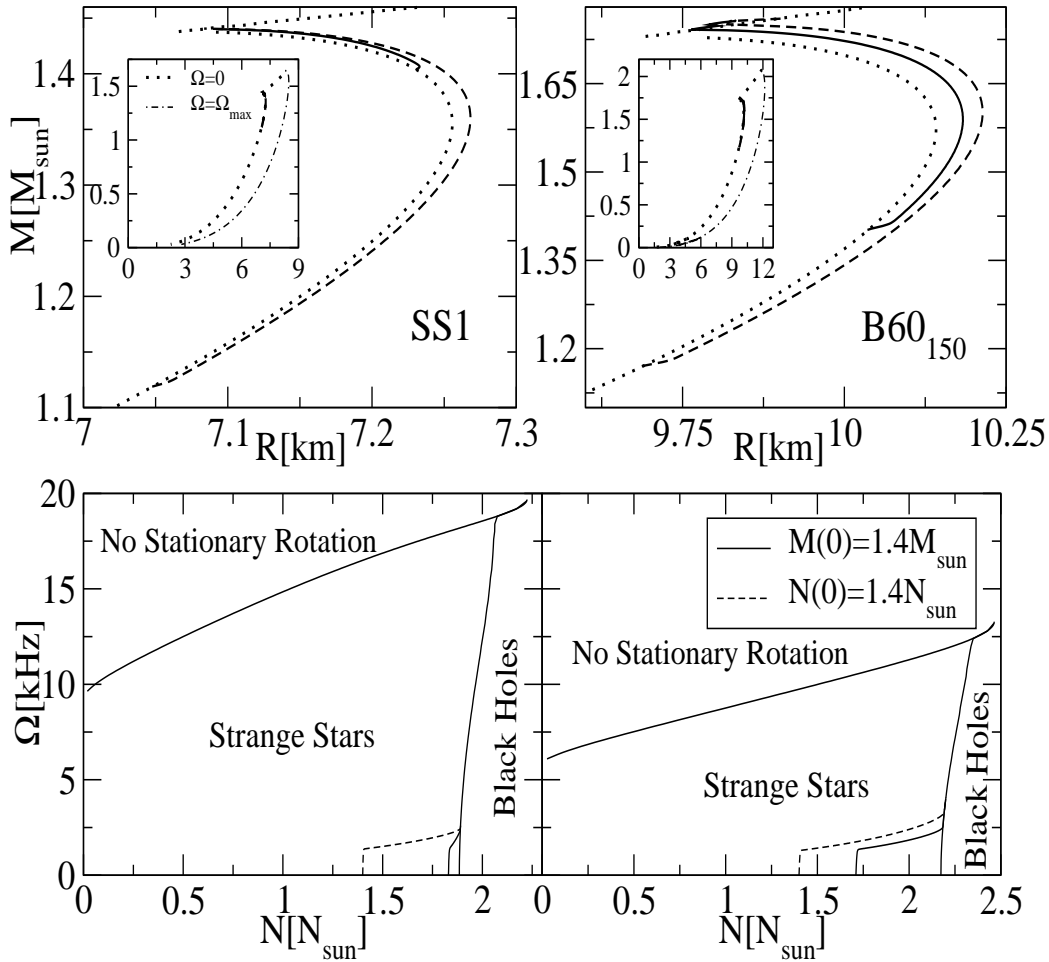,width=0.74\textheight,height=\textwidth,angle=-90}
\caption{Evolutionary paths for strange stars in the mass-radius
plane (top panels) and in the frequency-baryon number plane (lower
panels) for the equations of state SS1 (left panels) and
B60$_{150}$ (right panels), see text. } \label{fig:grphs}
\end{figure}

In Fig.~\ref{fig:grphs} we show evolutionary paths of accreting
strange stars in the mass-radius (MR) plane (top panels)
corresponding phase diagrams (lower panels) for two different EsoS
(SS1 right panels and B60$_{150}$ left panels). In each panel we
show two trajectories of a strange star initially rotating with
frequency $\Omega(0)= 0.001$ Hz; for which the initial magnetic field is
$B(0) = 7$ TG and its decay time is $\tau_B = 10^8$
yr. The mass accretion rate onto the star is $\dot M = 10^{-9}M_\odot
$/yr. The solid lines show the evolution of strange star configurations
with initial gravitational mass $M(0)= 1.4 M_\odot$, the dashed lines
show that of configurations with initial baryon mass $N(0)= 1.4 N_\odot$. 
In the insets of Fig.~\ref{fig:grphs} we show the stable branches for configurations rotating with maximal
frequency (dash-dotted lines) in compare to static ones (dotted lines).

From the evolution paths $\Omega(N)$ one can see that in all cases
of initial masses the magnetic braking force is strong enough to
stop fast spin-up of the star and saturate the frequency of
rotation. This can lead to an effect of frequency clustering.
However such effect for strange stars is correlated with the accretion
rate. 

\begin{figure}[htb]
\psfig{figure=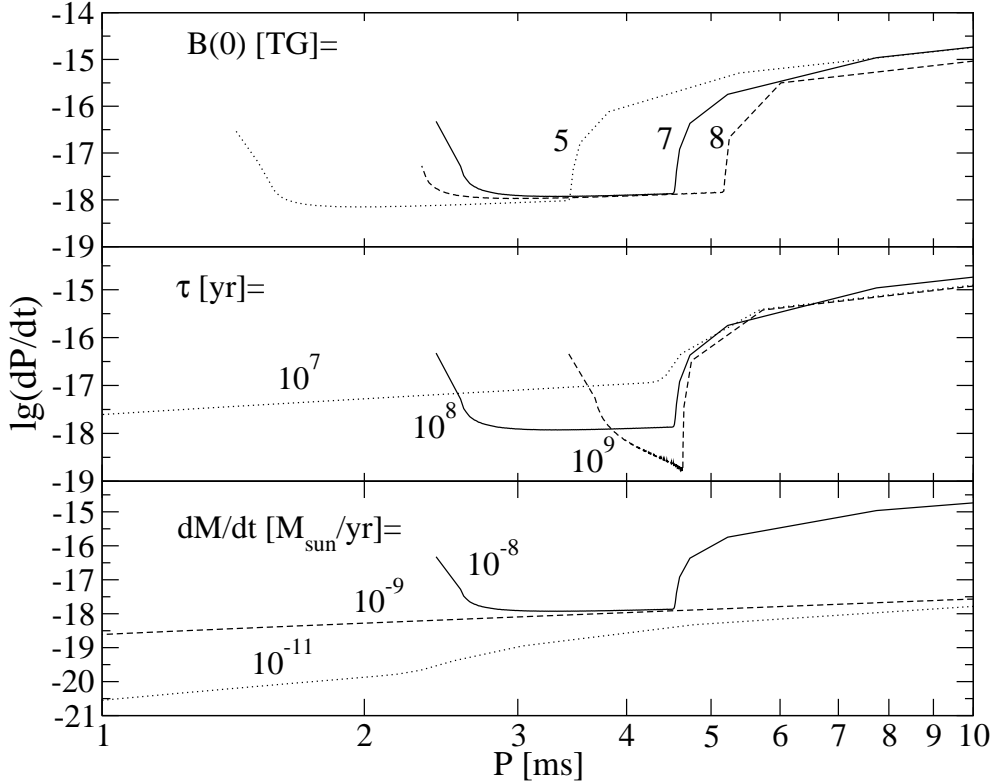,height=0.8\textwidth,angle=-90}
\caption{$\dot P$ - $P$ diagrams for accreting strange stars with
equation of state SS1. We show the dependence on the initial magnetic
field $B(0)$ (upper panel), 
mass accretion rate $\dot M$ (lower panel), 
and magnetic field decay time $\tau_B$ (middle panel).
}
\label{comper}
\end{figure}

In Fig. \ref{comper}, we show the result of our calculation for the spin
evolution of the accreting strange star with EoS SS1 in the $\dot P-P$
diagram, where $P=2\pi/\Omega$ is the period of rotation. 
The parameters of the accretion model are chosen such as to correspond to
values extracted from observations made on LMXBs, which are divided
into Z sources with $\dot M \sim 10^{-8} M_\odot/$yr and A(toll)
sources with $\dot M \sim 10^{-10} M_\odot/$yr
\citep{heuvel,klis,gw00}.

In Fig.~\ref{comper} we explore the sensitivity of the model to changes 
of the parameters with respect to the set used in Fig.~\ref{fig:grphs} 
($B(0)=7$ TG, $\tau_B = 10^8$ yr, $\dot M = 10^{-8} M_\odot /$yr) for
accretors with initial baryon mass $N(0)=1.4~N_\odot$. 
In the upper panel we vary the inital magnetic field
$B(0)= 5, 7, 8$ TG, in the middle one the magnetic field decay time
$\tau_B = 10^7, 10^8, 10^9$ yr and in the lower panel the accretion 
rate~ $\dot M = 10^{-11}, 10^{-9}, 10^{-8} M_\odot /$yr.

We see from this figure that changes of the
accretion parameters not only shift (choice of the initial magnetic field)
and deform (choice of the magnetic field decay time) the
interval where the spin-up is saturated (dip in $\dot P$), but
for some cases this effect can be
washed out by a variation of the accretion rate. 
This phenomenon means that the existence of a frequency clustering for
strange stars requires a strong limitation of possible values of the
accretion parameters.

\begin{figure}[ht]
\psfig{figure=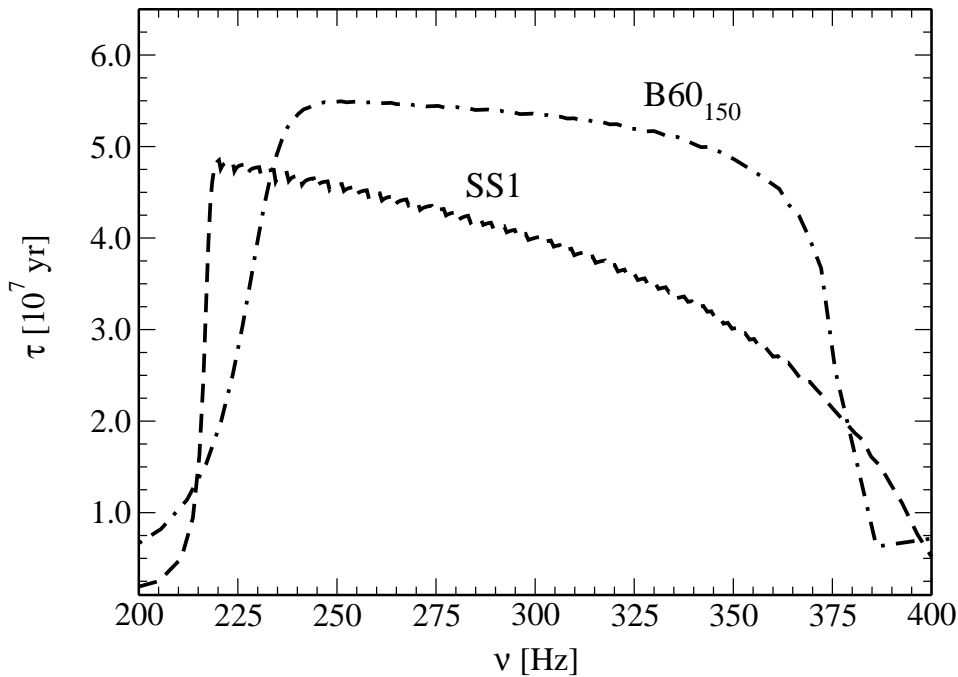,height=0.8\textwidth,angle=-90}
\caption{Waiting times as a function of spin frequency for spin-up
evolution of strange stars with equations of state B60$_{150}$ and
SS1.} \label{waitss}
\end{figure}

In Fig.~\ref{waitss} we plot the 'Waiting time' $\tau=|P/\dot P|$ 
\citep{accmag}
of the strange star as a function of the spin 
frequency.
for the EsoS SS1 and B60$_{150}$.
In order to obtain an enhanced waiting time (frequency clustering) in the 
interval $220$ Hz $<\nu< 380$ Hz which corresponds to recent observations
\citep{gw00},
we have to choose the following parameters:
$B(0)= 2.5$TG, $\tau_B = 10^8$ yr, $\dot M = 10^{-9} M_\odot /$yr
for SS1 model and $B(0)= 3$ TG, $\tau_B = 10^8$ yr, $\dot M =
10^{-8} M_\odot /$yr for B60$_{150}$ model. For
both cases the initial gravitational mass is the same $1.4
M_\odot$, which corresponds to $1.83 N_\odot$ initial baryon mass
for SS1 and $1.71 N_\odot$ for B60$_{150}$ model. 
The main difference between both of these scenarios for the frequency
clustering is the initial baryon mass and the mass accretion rate.
An independent determination of these quantities to a sufficient accuracy 
could thus rule out one of the compact star models.
At present, this could be done only for a minority of objects 
\citep{LM00}
and bears some model dependence.
\citep{accmag}.

\section{Conclusions}
We have investigated the question whether the clustering of spin frequencies
which has been observed for the compact objects in LMXBs is 
consistent with the hypothesis that at least some of these objects are 
strange stars.
We have applied a standard model of magnetic disc accretion
and find that population clustering in a narrow band of spin frequencies 
between $220 \le \nu[{\rm Hz}] \le 380$ can occur for typical parameter values 
of the model.
Inspection of the response to parameter variations shows that the lowering of
the magnetic field decay time can wash out the effect as well as a change in
the accretion rate. The changes in the initial magnetic field leave the 
waiting time distribution rather unchanged but shift the interval of period 
clustering.

On the other hand, the observation of frequency clustering alone is no 
indication for the presence of a strange star.  
A standard bag model EoS shows a similar waiting time pattern as the strange
quark matter EoS does.
For the EsoS investigated in this paper no clustering of masses has been 
obtained. This is a striking difference to hybrid star configurations
\citep{accmag}, 
which could be used 
in order to constrain further our approaches to the EoS of superdense matter
provided a sufficiently large sample of LMXBs could be observed and their
frequencies and masses could both be extracted with sufficient accuracy.  

\section*{Acknowledgement}
I.B., H.G. and G.P. acknowledge the hospitality of Rostock University.
This work was supported in part by the {\sc Deutsche
Forschungsgemeinschaft} (DFG) under Grants No.  436 ARM 17/7/00,
436 ARM 17/5/01, 
by the {\sc Graduiertenkolleg} ``Stark korrelierte Vielteilchensysteme'' and
by the {\sc Deutscher Akademischer Austauschdienst (DAAD)}.
We thank D. Aguilera for discussions and careful reading of the
manuscript.

\end{document}